# Inferring Networked Device Categories from Low-Level Activity Indicators

Kyumars Sheykh Esmaili
Nokia Bell Labs
Antwerp, Belgium

Jaideep Chandrashekar
Technicolor R&I
Los Altos, CA

Pascal Le Guyadec
Technicolor R&I
Rennes, France

## ABSTRACT

We study the problem of inferring the *type* of a networked device in a home network by leveraging low level traffic activity indicators seen at commodity home gateways. We analyze a dataset of detailed device network activity obtained from 240 subscriber homes of a large European ISP and extract a number of traffic and spatial fingerprints for individual devices. We develop a two level taxonomy to describe devices onto which we map individual devices using a number of heuristics. We leverage the heuristically derived labels to train classifiers that distinguish device classes based on the traffic and spatial fingerprints of a device. Our results show an accuracy level up to 91% for the coarse level category and up to 84% for the fine grained category. By incorporating information from other sources (e.g., MAC OUI), we are able to further improve accuracy to above 97% and 92%, respectively. Finally, we also extract a set of simple and human-readable rules that concisely capture the behavior of these distinct device categories.

## 1. INTRODUCTION

To a home Internet gateway (and the ISP operating the gateway), devices on a home network are represented by a MAC address (and perhaps a hostname). Associating a higher level semantic to a device (*smartphone*, *game console*, *smart meter*) to devices benefits a wide range of scenarios. For example, this capability will allow a ISP help-desk operator to quickly identify the particular device (with MAC address) that a customer is complaining about; an ISP could better target particular broadband or other service offerings to subscribers by profiling the devices used; an ISP subscriber can set-up traffic policies for their home network more expressively (prioritize traffic from *game consoles*), and so on. Such a capability is particularly relevant today given the proliferation of networked devices that are growing the Internet-of-Things (weighing scales, smart meters, thermostats, etc.). In this paper, we investigate device classification from the point of view of a home Internet gateway (and the ISP). To this end, we analyze a large dataset of device activity, extract a set of features that allow us to differentiate between different types of devices (and to different extents) and apply known classification techniques to map individual devices to device categories.

While there are a number of different approaches to identifying device types, the vantage point (the home gateway) and the interested entity (the ISP) present a number of challenges and restrict the space of solutions. A method that relies on packet inspection could reliably uncover the device type (e.g., by examining HTTP U/A strings). However, deep packet inspection is not viable on most of today's resource constrained home gateways. While ISPs could easily implement such a function upstream, they are very reluctant to do so due to consumer privacy concerns [5]. Yet another method leverages the fact that user's often assign descriptive names to their devices (e.g., `John-iPhone`) which can be parsed to infer the device type. Unfortunately, relatively *few* devices in the home involve the configuration step (typically personal devices) and the majority have default strings which reveal little about the nature of the device (we present data to support this later in the paper). Finally, one may look to exploit information in the MAC OUI (which identifies the device manufacturer). However, the discriminative power of the MAC OUI depends greatly on the product range of a manufacturer and the incidence of a manufacturer in a particular dataset. We explore this issue in detail in a later section.

In this paper, we develop a systematic methodology for accurately classifying devices based on their *behavior* while on the home network. Our methodology is based on measurements and metrics that are already available today in commodity gateways deployed by all ISPs. The underlying intuition is that the devices of a certain class – say, tablets – are likely to be used by their owners in similar ways and we expect their long term behavior to be more similar to each other than devices from a different category, for instance – laptops. For example network extenders (which aggregate several devices behind them) are likely to have higher traffic volumes than mobile devices or peripheral devices (e.g., printers).

Based on this intuition, we extract behavioral signatures that capture a device's behavior along several dimensions: *how much traffic it generates, how much it moves, what is its daily usage?* The signatures that we construct for a device leverage common wisdom captured in previous work; we also develop a number of novel behavior traits that are shown to have good discriminative power. We develop a two level taxonomy of device categories – a coarse grained device class and a finer grained device type – based on an informal survey of home network users. We then cast the problem of identifying the device categor(ies) as a multi-class classification problem and present the results from applying two well known classification methods – Decision Trees and Support Vector Machines. While our dataset contains both wired and wireless devices, the work is this paper mainly focuses on wireless devices which constitute the majority of the device population. However, our methodology is general and may be applied across the wider set of devices. The key contributions of this paper are summarized as follows:

1. We explore in detail a large dataset of home network activity taken from over 240 subscribers of an ISP for a period of 1 year and which contains observations of over 5000 devices. We then identify and extract a number of features, a number of which are identified based on insights obtained from the data exploration, which capture traffic and spatial behavior of individual devices.

2. We develop a two level taxonomy of common home network devices that associates a device with a coarse grained device *class* and a finer grained device *type*. We then use a number of simple heuristics over the static descriptors for a device to map each device onto the two categories. We are able to obtain coarse grained labels for 62% of the devices and fine-grained labels for 36% of all devices in our dataset.

3. We study the performance of two well established classification techniques – Decision trees and SVM – to infer the labels for 1878 wirelessly connected devices. We find that the SVM based classifier provides the best classification accuracy and is 90.47% accurate on coarse grained labels (83.11% for fine grained labels). Furthermore, we explore the impact of incorporating other information such as the MAC OUI and we show an additional improvement of roughly 8-10% for the fine grained labels, and $\tilde{6}$% for the coarse grained labels. We further examine the Decision Tree generated to undersatnd the importance of particular features that influence the type of a device.

The rest of this paper is structured as follows. In §2, we briefly describe related work differentiate them from our contributions. In §3, we describe the device dataset used in this paper and in §4, we present our device taxonomy and describe some heuristics that we apply to extract labels for a subset of the devices. We discuss feature extraction in detail in §5 and explain the intuition behind each of features extracted from the raw metrics. In §6, we introduce two classification methods and report on the results of applying these on our dataset(s). in §7, we conclude by discussing some open issues and extensions.

## 2. RELATED WORK

In this paper, we study the problem of identifying the *type* or *class* of a networked device, and disambiguating between devices of different classes coexisting on the same home network. While we are not aware of any work in the literature that looks at this specific problem, there has been considerable work in closely related areas – Traffic & Application Inference, Device Disambiguation and Device Fingerprinting. We briefly discuss representative work in each.

**Traffic & Application Inference:** A large body of work has addressed the problem of determining the underlying applications corresponding to traffic flows observed in the network. The high level intuition behind all of these is that the differences in application classes is reflected in the composition of traffic and packets generated by these applications. For e.g., the graphical relationships between source(s) and destinations(s) are exploited to identify traffic classes, as is described [14]. In another approach, an initial inference is improved upon by incorporating more global information [12]. A comprehensive survey of the techniques in this area can be found in [20] and we highlight two other notable works. In contrast to all of these, our work focuses on identifying *device classes*.

**Device Disambiguation:** A number of previous efforts have addressed the problem of distinguishing devices behind a firewall or NAT device. For example, Bellovin proposed a method to exploit IP header fields to identify distinct hosts sharing a single IP address [2]. In [18], the authors describe an approach relying on the IP TTL field and HTTP user-agent strings to correctly count and distinguish devices behind a single DSL line. While these are applicable to a narrow part of our work (identifying devices behind a non-transparent network bridge, for instance), the data available to us is less semantically rich (aggregated byte counters, in our case) and thus these methods are not directly applicable.

**Device Fingerprinting:** There are two main approaches to uniquely identifying devices (or software stacks, for that matter). In the *active* approach, a set of crafted packets is directed at a device/network interface and the responses are analyzed to identify particular customizations or properties of the underlying code base – these constitute a signature of the entity being fingerprinted. This is the approach taken with some security scanners (e.g., Nmap [21]), browser fingerprinting tools [7], or wireless chipset [3]. In contrast, *passive* fingerprinting methods attempt to extract unique signatures from simply observing the device interacting with the network.

This approach has been applied very successfully in the past in fingerprinting wireless cards and chipsets [9, 22], or even devices [16]. Our own work differs from these in that rather than extract unique fingerprints of a particular device or chipset, we seek to extract (behavioral) signatures for a certain device type (which is greatly affected by when, where and how often a device is used). While some of the cited methods may be applicable in general, our specific approach is dictated by the specific nature of our dataset (which does not support any of the other methods).

## 3. DATASET

Our work in this paper uses a dataset of over 5789 devices collected across 240 distinct homes over a year (from 01/11/2013 to 01/10/2014). Here, we elaborate how this dataset was collected and describe the recorded metrics in detail. The 240 households captured in our dataset were subscribers of a (single) large European ISP during the period indicated and were recruited to take part in a trial deployment. Home gateways in these households were identical to all other subscribers except in one aspect: gateways in the deployment periodically queried a set of software counters and state variables on the gateway and reported these to a back-end server. The subscriber home gateways, spanning 4 distinct hardware models, were configured and provisioned for a triple-play service (voice, broadband, IPTV) and came with 4 physical Ethernet ports and a 802.11b/g/n 2x2 WiFi access point. Importantly, the WiFi access point of every gateway in the trial was operational and in use for the entire period.

### 3.1 Collected Raw Metrics

In the following, we enumerate and describe in detail the particular set of recorded parameters that are relevant to our work in this paper. We note that each gateway observes *all* of the devices that were on the home network at any point during the deployment and thus our dataset is a complete record of all the network activity across the 240 households during the period. If there are any mobile cellular devices that are not configured to use the home WiFi network (perhaps exclusively using the cellular network), there is no record of it in our dataset.

**Host Descriptors:** These correspond to the *static* properties of individual devices that do not change over time and include the following:

•*MAC Address:* the 48 bit address of the network interface that is connected to the local area network. The first three octets of the MAC address (the `OUI`) identifies the vendor of the network interface. In most cases – but not all – this also corresponds to the device manufacturer.

•*HostName:* a descriptive string associated with a particular device, as recorded in the DHCP table of the home gateway. In the case of many personal devices – computers, phones, tablets, etc., – users often select descriptive, parse-able strings that can hint at the type of device (and ownership). However, in the more common case, either the device self-selects a hostname or else the gateway assigns a default hostname that incorporates the MAC address (to ensure it is unique).

•*Physical Interface:* describes the nature of the connectivity to the gateway and is either wired (via one of the 4 Ethernet ports on the gateway) or over WiFi.

The host descriptions do not change often and are reported infrequently (every hour or so). In contrast the dynamic variables and parameters that we discuss next can change very frequently and are reported more frequently by the gateway (roughly every minute).

**Traffic Volume Metrics:** These report cumulative traffic volumes (in bytes) associated with particular (physical) gateway interfaces (for wired devices), or stations (for wireless devices). We can compute the traffic seen between samples by computing successive differences. In the case of WiFi stations, traffic records are indexed by the MAC address of the station. We point out an important artifact that is observed in our dataset. The deployment covers 4 distinct traffic models, and in the case of all but one, the wired traffic counters use 16 bits (and wrap around even with light traffic). In these cases, we are unable to accurately extract the traffic volumes associated with the devices connected on those ports. While we do have some gateways that use 32 bit counters, they are few in number. Due to the paucity of traffic data on the wired ports, we remove wired devices from consideration and focus mainly on classifying wireless devices – which are always associated with 32 bit counters.

**Wireless Metrics:** For the wireless station that is associated with the gateway, two additional metrics are reported:

• *Traffic Rates:* these are reports of the actual throughput (sent and received bits per unit time) experienced by each station associated with the gateway AP. Importantly, the gateway driver represents these as integer values that are clipped to zero when the actual value is less than 1 *kbps*. Note that this rate accounts for low level management overhead on the wireless network (not accounted for in the traffic volume metric).

•*RSSI:* This captures the signal strength from the *to the gateway* to the station. High(er) values typically represent better coverage. Per station RSSI values are computed at the gateway based on received data frames. The actual value reported by the gateway is an average value over all received packets in the reporting interval.

# 4. DEVICE TAXONOMY & LABELING

We polled 140 distinct subscriber households from our dataset and asked them to enumerate all the various networked devices in their home (along with the corresponding MAC address). We received results from 34 homes, which together named 137 devices. Table 2 presents a very small selection of these. The hostnames indicated in the first column were obtained by checking the MAC addresses against the DHCP entries in the dataset. The last column reflects the user description of the device. Based on our intuition from analyzing this data, we constructed a two level taxonomy of device categories. The lower level – the fine-grained class – captures how *most* users' identify their devices. At a higher level, we also categorize the devices into coarse grained classes which model how users relate to their devices, how they use them, and how these devices behave on the network. We utilize the coarse grained labels for two reasons. First, we see a gradual blending of device functions as device form factors converge. For example, the so called "phablet" form factors blur the line between smartphones and tablets. Second, relying purely on the fine grained classification results in a very unbalanced data and poor results for under-represented devices (in our dataset). Using a two level taxonomy addresses both of these factors and this categorization is presented in Table 1.

The first row enumerates the coarse grained classes. *Compute* devices are essentially laptops or desktops, which are general purpose and used for a variety of tasks by users (and sometimes shared between users). *Mobile Handheld devices* are devices typically associated with an individual and often carried on their person (or carried around a home) and where the interaction periods are shorter (than compute devices). *Network Equipment* refers to devices that extend the range or functionality of a device, and typically aggregate the behavior of several other devices placed behind them. Finally, we have the "Consumer Electronics" (*CE*) category of more specialized devices, which are built or optimized for specific tasks or to consume particular services.

The second row enumerates all the fine-grained classes inside of each coarse-grained category For example smartphones, tablets and eBook Readers all come under under the high level categorization of *Mobile Handheld* devices. With respect to the *Compute Device* class, we do not differentiate desktops and laptop computers. In general, we do not expect significant differences in how people *use* these devices. While a laptop may be used in more than one location in a home (while a desktop generally stays in a single location); we expect they are both used the same way (as a general purpose compute device) to do different things (check email, browse the web, etc). Pragmatically, distinguishing these two is that extracting ground truth labels are difficult; we explain this further in the next section. Finally, we have devices such as printers/scanners, ISP supplied set-top-boxes (STB), over-the-top (OTT) devices, and media or game consoles, which are often designed to support one particular use case. We point out that this categorization is slightly arbitrary - several game consoles today support video streaming (as do OTT boxes). This does complicate the inference of the fine grained labels, the coarse grained label inference is less sensitive. We also note the fine grained categorization is not exhaustive – we observe single instances of several device types in our dataset (a smartwatch, a wireless smart-meter, etc.). We ignore device types for which we observe less than 3 instances in our dataset. In the following, we describe a number of heuristics to infer the coarse and fine grained labels for a device (as identified by its MAC address); these heuristics are based on our intuition and a careful exploration of the dataset.

## 4.1 Extracting "Ground Truth"

We developed a set of heuristics against the *Host Descriptors* for all the devices in our dataset and this yielded a smaller set of devices for which we believe the inferred labels are accurate. We additionally verified the heuristics against the 137 devices for which user supplied labels and ensured that the heuristic inferred labels are consistent with that supplied by users.

•**Name Based Heuristics:** Hostnames often contain common descriptive strings, e.g., android, pc, iMac, etc., which can be leveraged to infer the device's category. Often, the descriptive strings are specific enough to yield a finer grained categorization (e.g., `john`-iphone), in which case we have both a coarse label (mobile handheld) as well as a fine grained label (smartphone). Sometimes, hostnames reflect very specific device models (e.g., `wrt54g` – a home router manufactured by `Linksys`) or take on factory default names that are well known (e.g., the ISP supplied STB is named identically across all the homes it is deployed in). In such cases, we are able to infer the nature of the device.

•**MAC Address Refinement:** In some cases, the MAC address of a device can be helpful in its identification, particularly when the vendor (identified by the MAC OUI) manufactures has a very limited product range (that may live in a home). One example: the MAC OUI `00:00:48` is registered to `Seiko Epson` corporation; it is *likely* that devices with this prefix are printers or scanners. However, the OUI is far less useful in distinguishing devices in the case of vendors with diverse product portfolios. For instance, Apple Inc. manufactures devices that span all of the coarse grained categories in Table 1. Here, knowing that a device is manufactured by Apple does not allow us to identify the particular category. Yet another caveat in using the MAC address is that it may be incorrect (on rare occasion): we saw one corner case where a vendor incorrectly "borrowed" the mac OUI of a different manufacturer (this was also reported in [17]).

| Compute Device | Mobile Hand-Held | Networking Equipment | CE |
|---|---|---|---|
| Laptop/Desktop | Smartphone Tablet e-Reader | PowLine Eth. WiFi Ext | Smart TV NAS Game Console Media Bridge OTT box Printer/Scanner |

Table 1: Device Categories

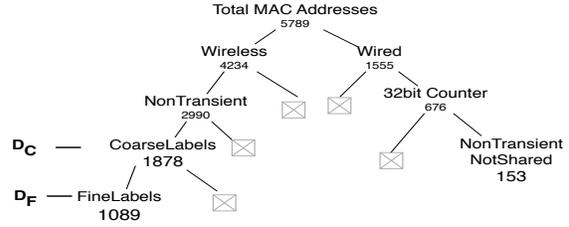

Figure 1: Overall Dataset Summary

•**BSSID based:** Several printers and media bridge devices allow WiFi-direct connections for which the device temporarily advertises a network that uses the device's MAC address (or an adjacent address) as the BSSID. In some cases, the advertised SSIDs explicitly contain the device model we are able to leverage these. We are able to identify some wireless range extenders or network bridges using the same technique.

•**Other heuristics:** We identified several instances in the *Host Descriptor* data where a single MAC address is associated with a number of hostnames (which are unrelated), and this occurs when several (potentially!) devices are connected *behind* the network extender device. When this count is greater than two, we tentatively label the device as a network extender. Then, we carry out an additional check to verify if there are overlapping sessions (distinct MAC addresses associated with the same name sending/receiving traffic at the same time). If such behavior is found, we confirm that the device is a network extender. We distinguish between Powerline Ethernet connectors and WiFi Extenders by comparing the set of hostnames against an exhaustive set of models of each, and also leveraging the MAC OUI.

Finally, we present a census of all the devices that are present in our dataset in Figure 1. We see host descriptors for 5789 distinct mac addresses. Of these, 4234 are associated with the wireless network, and the rest are wired devices. Of the 1555 wired devices, only 676 are observed on the gateways that use 32 bit traffic counters. Further, of the remaining, only 153 devices were found not to share the physical port with other devices (which is essential to correctly attributing traffic to the device). Given this very small population (relative to the whole), we omit wired devices from consideration in the rest of this paper. Moving now to the devices seen on the wireless network, we find 1244 devices to be transient; that is, they are observed for very brief periods of time on the network- the duration being insufficient to contribute sufficient metrics to aid in classification (this is further discussed in the data sufficiency experiments in Section 6); we remove these from consideration. Finally, we are left with 2990 *non-transient* wireless devices for which sufficient data is available. However, we can establish the ground truth categorization only in 1878 instances for the coarse categories (denoted $D_C$), and in 1089 cases with a fine grained label (denoted $D_F$). Unless indicated otherwise, the results and analysis in the paper pertains to the sets $D_C$ and $D_F$. It should be pointed out that our labeling is conservative. We also implemented a number of sanity checks, for e.g., a tablet or a smartphone cannot appear on a wired interface without an extender present, an android phone could not be manufactured by `Apple`, and so on). Any discordance between the inferred labels and the sanity check led to the label being rejected. By being conservative, we are unable to categorize a large number of devices in our dataset; however, we have a high degree of confidence in the extracted labels are accurate.

It should be clearly pointed out that the labeling that we described in this section is only to obtain training and testing data for the statistical classification methods (Section 6.1). It is unlikely that an ISP would use these heuristics by themselves and this for two reasons: (i) the heuristics described are unable to classify a large number of devices, and (ii) there are potential privacy concerns with an ISP being able to read hostname strings, which can sometimes contain names of people.

## 5. EXTRACTING FEATURES

In this section, we analyze the time series' of traffic and wireless metrics and subsequently construct a number of features that are suitable to input to a classification framework. We use the term *feature class* to describe a certain metric, possibly post processed, collected from the gateways and from which a distribution is constructed, and *feature* to refer to a specific attribute extracted for that feature class. As an example, *Daily Traffic Volume* is a timeseries and a feature class, while the median daily traffic volume is a feature. In the following, we walk through each feature class and describe how they are processed. In selecting these features, we rely on a combination of common best practices and also intuition obtained from initial explorations of the dataset.

The complete set of 92 features that are associated with each individual device is summarized in Table 3. Table Table 3(a) enumerates feature classes while 3(b) represents the features extracted from each class. Finally, Table 3(c) describes 4 additional features that are all extracted from the RSSI feature class.

|   | Hostnames | Connecting Interface | Category | Sub-Category |
|---|---|---|---|---|
| 1 | `user-iPhone` | WiFi | Mobile hand-held | Smartphone |
| 2 | `iPad-user` | WiFi | Mobile hand-held | Tablet |
| 3 | `user-toshiba-laptop` | Wired | Compute | Desktop/Laptop |
| 4 | `PC` | Wifi | Compute | Desktop/Laptop |
| 5 | `TL-WA850RE` | Wired | Network Eqpmt. | WiFi Repeater |
| 6 | `android-2013051200001053` | WiFi | Mobile hand-held | unknown |
| 7 | `RM4100` | Wired | Consumer | STB |

Table 2: Sample hostnames and fine grained labels obtained from survey (and the assigned coarse labels). The device listed on row 6 was named "mobile" by the subscribers, which we consider a coarse grained label. Finally, row 7 reflects a well known naming convention used by the ISP for set top boxes.

| Feature Class | Subclasses |
|---|---|
| Daily Traffic Volume | `tx` |
|  | `rx` |
| Session Length | `session_l` |
| Per Session Traffic Volume | `session_tx` |
|  | `session_rx` |
| Traffic Rate | `tx_rate` |
|  | `rx_rate` |
| RSSI | `rssi` |

(a) Time series features

```
count
min
p10
p25 (Q1)
p50 (median)
p75 (Q3)
p90
max
p75-p25 (IQR)
p80-p20 (range80)
max-min (diameter)
```

(b) Different point-summaries (11 in total) used for each time series

| Feature Name | notation |
|---|---|
| Allan Deviation | `rssi_ad` |
| M·D product | `rssi_md` |
| # Locations | `rssi_num_locations` |
| AuC | `rssi_auc` |

(c) Single-Point Features `AuC` indicates the area-under-curve corresponding to the empirical density function.

Table 3: All extracted features: (a) the feature classes/subclasses that are in the form of time series, (b) the 11 statistical point-summaries used for each time series, (c) the single point (i.e., non-time series) features

## 5.1 Feature Classes

We first describe all the feature classes that are constructed from traffic (volume and rate) counters, and subsequently describe the feature classes corresponding to the RSSI data.

**Traffic-based Features**

*Daily Traffic Volume:* The amount of data (in bytes) that a device transmits (or receives) each day that it was active. Typically, we expect devices such as PCs, OTT video streamers, or network extenders (which aggregate traffic for others) to have higher values for this feature (in contrast to mobile hand-held devices).

Importantly, we also use this feature to distinguish between transient and non-transient devices: a device is non-transient if it has non-zero traffic volume for at least 3 days (setting this threshold is discussed in the Results section) and transient otherwise[1]. The rest of this paper only considers devices that are non-transient.

Figure 2a depicts the median of medians of outgoing (from the gateway to the device) daily traffic for the devices within each group (the incoming traffic graphs exhibit similar patterns). It shows that PCs and Network Extenders consume the largest amount traffic per days. There is also a considerable difference between Tablets and Smartphones.

*Session Length:* The duration of time that a device was active, i.e., generating traffic, is computed as follows: sessions are initiated when the corresponding traffic counter is non-zero and terminated when there is no traffic activity for at least 15 minutes (this is the ARP timeout value). Thus, for each device, we obtain a list of sessions of the form (start time, end time). Intuitively, we expect small hand-held devices to have shorter sessions, but more of them, than fixed devices such as computers and set top boxes. The result is a time series of session lengths for each device and from this, we construct a distribution. Figure 2b depicts the distribution of *median* session length across the various (coarse) device classes. Here, the distribution for mobile devices (smartphones, tablets) reflects shorter sessions than for devices like OTT boxes, Network Extenders, or even PCs (slightly).

*Per Session Traffic:* is the cumulative traffic volume associated with each distinct session for a device. To allow for fair comparisons across different sessions, we normalize these volumes by the length of the session. We then construct a distribution for each device with these normalized values and extract all the previously listed statistics as individual features. Figure 2c shows the distribution of the median values for the normalized session rates.

*Traffic Rate:* We build a distribution over the traffic rate metrics obtained for wireless devices. Recall from Section 3.1 that this rate only accounts for user traffic (ignoring the effect of wireless control traffic) and thus contains a large number of zeros. We find that the median traffic rate across almost all the devices in our dataset is zero. In fact, we expect this particular fact to help discriminate between device types – we expect devices such as computers to have short(er) runs of zero values (generally when they're on, they are used) as compared to mobile devices, where the device can be on for long periods of time with only some minimal activities. Similarly, we expect fewer zeros associated with network extenders (which aggregate traffic from other devices).

---
[1] These could, for example, be visiting guests' devices that have been connected to the home's AP.

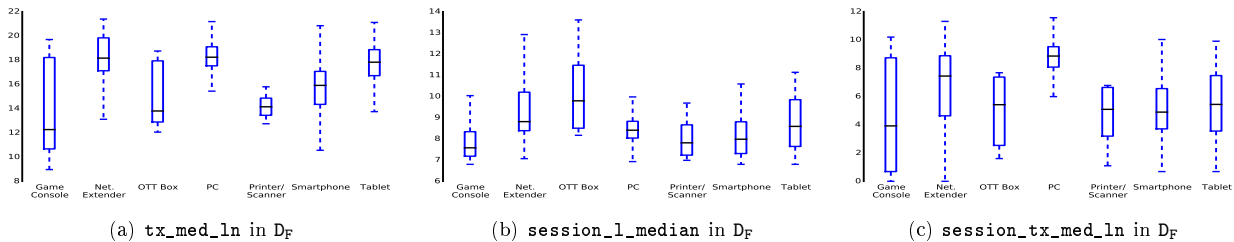

Figure 2: Samples of Traffic based features across fine grained device classes

**Spatial and Mobility Features**

The RSSI reported for each device (as recorded by the gateway) loosely tracks the physical distance of the device from the gateway [13,15,19]. For a given network environment (and device), we expect that distances increase as the RSSI increases. Thus, as a device is used in different locations over time, the set of all RSSI measurements constitutes a spatial signature and the short term differences in the RSSI captures the mobility pattern of the device inside the home. With this intuition, we extract a number of feature classes to capture this aspect of the device behavior.

*Raw RSSI:* For each device, we construct a distribution of the reported values based on the timeseries of the raw RSSI values. From this, we extract all the typical distribution related metrics as we do for the traffic features (max, median, iqr, etc). Note that we include the minimum value of the distribution as a feature, since it is not uniformly zero (across devices), unlike in the traffic scenario.

*RSSI Diameter:* This is computed as the difference of the extreme values in the RSSI distribution, i.e., the max-min statistic. Note that this is not as "useful" for the traffic based features where the minimum values across devices are generally zero. Figure 3a shows the distribution of this feature across the coarse grained device classes. From the figure, the distribution for mobile hand held devices is shifted slightly higher than the other classes; this indicates that such devices have higher portability, i.e., can be used in a wider spatial range. *RSSI M-D product:* This feature is constructed as the product of the RSSI Diameter (previous) with the median value of the RSSI distribution. We include this feature to amplify the differences (in the individual product terms) across the device classes.

*RSSI Allan Deviation:* Specific to the timeseries of RSSI values $X = \{x_1, x_2, \ldots, x_n\}$, this is defined as

$$\mathtt{AD}(X) = \sqrt{\frac{1}{2(n-1)} \sum_{i=1}^{n-1} (x_{i+1} - x_i)^2}$$

Allan Deviation (or AD, for short) captures the instantaneous variation for time-indexed data and was originally designed to study frequency stability in clocked analog circuits [1]. Applied to the RSSI time series, the allan deviation captures the short time scale mobility of a device. We expect this to be high(er) for mobile hand held devices as compared to laptops or other more stationary devices. Fig. 4 shows the distribution of this feature over the different device classes. In Fig. 4a, we see that the distribution for mobile hand-helds is pushed upwards; this is expected, since we expect these devices to be carried by users as they move around and this is captured in short-term variations in the observed RSSI. In Fig. 4b, the distributions for smartphones and tablets are quite different, with the former tending to have higher AD values. Again, this is borne out by intuition; a smartphone might be in a user's pocket and connected to the network as they walk around the home.

*Location Modes:* This feature quantifies the number of *modes* in the rssi distribution. Intuitively, it captures the number of locations inside a home, where the device is used most. By exploring the data, we find that mobile devices have *wider* distributions (reflecting usage at multiple locations and in-motion), while on the other hand static devices (desktops) have narrow RSSI distributions and single modes. To extract this feature, we developed a segmentation algorithm, inspired by the Median Shift algorithm [24] which infers the number of local maxima in a distribution, to identify the modes of the RSSI distribution. We omit a detailed description of the algorithm due to a lack of space.

This metric is very intuitive and ought to clearly distinguish device types used in multiple places from those that are more fixed. However, the number of locations reported is quite low across all devices (at most 3). Upon close examination, we find a number of mobile phones to have single modes, but extremely wide distributions; we believe that natural variations and noise in the RSSI result in washing out the peaks. To deal with this, we introduce the next feature.

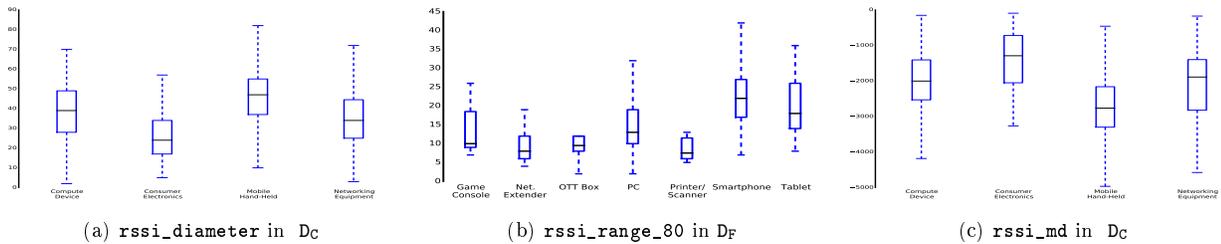

(a) `rssi_diameter` in $D_C$    (b) `rssi_range_80` in $D_F$    (c) `rssi_md` in $D_C$

Figure 3: Mobility Related Features

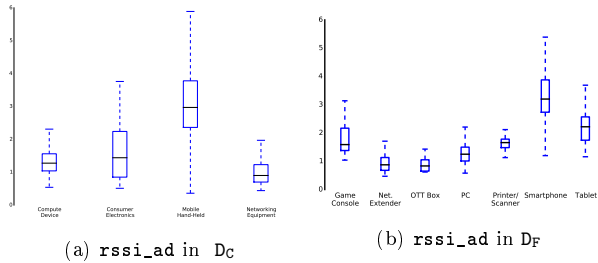

(a) `rssi_ad` in $D_C$    (b) `rssi_ad` in $D_F$

Figure 4: Distribution of Allan Deviation of RSSI for different device classes: coarse and fine grained categories

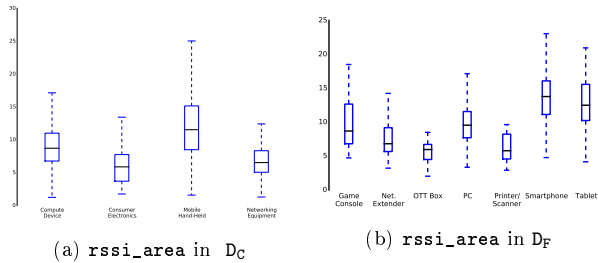

(a) `rssi_area` in $D_C$    (b) `rssi_area` in $D_F$

Figure 5: Distribution of RSSI AuC for different device classes: coarse and fine grained categories

*RSSI AuC:* This feature is computed as the area under the curve of the RSSI distribution. The main intuition here is that this area captures the degree of *in-motion usage*. In order to allow for fair comparisons across different devices (with different scales), we normalize the AuC value by dividing with the absolute maximum frequency. One immediate conclusion is that the higher the AUC is the more in-motion usage the device has. Figure 5 shows the distribution of this metric across the different device classes. This feature complements the Allan Deviation, as it is more resilient to noise. Concretely, a few noisy measurements can have a large value on the Allan Deviation as computed, especially for devices with short RSSI time series, the impact is less pronounced on the computed AuC values.

### 5.2 Feature Analysis

In this section we post-process the features extracted previously by applying some standard techniques. The goal here is to transform the features so as to improve their ability to discriminate between different classes, and thus improve the classification results.

#### 5.2.1 Feature Rescaling

Many statistical models, including the classification methods that we use in this paper, make implicit assumptions about the data being *Gaussian*; when this is not the case, classification performance may be improved by transforming them. We systematically explored the features to identify those with heavily-skewed distributions. We use the *Extreme Value* detector which proceeds by identifying the fraction of the population that lies outside a "central" range. Specifically, it labels all the values $x$, such that $Q_1 - EVF * IQR > x$ or $x > Q_3 + EVF * IQR$. Here $Q_1$ ($Q_3$) are the first (third) quartiles, and `IQR` is the inter-quartile range, i.e. $Q_3 - Q_1$. If at least 1% of the values in a particular feature lies outside this interval, we consider the feature to be excessively "skewed". Table 4 enumerates the 10 most skewed features extracted from our data for `EVF`=6 (the skewness test is fairly stable and we did not observe different outcomes with `EVF` in [3, 10]. It is worth noting that all of detected features are skewed only to the right side of the distributions. In short, the two feature families of our datasets – traffic based, and rssi based – behave differently. Distributions constructed from RSSI based features have a Gaussian shape. On the other hand, distributions from traffic based features tend to have long tails. This is likely because traffic phenomenon are inherently bursty, and devices often have long inactivity periods during which no traffic is generated.

Consequently, for each attribute identified as skewed, a logarithmic transformation is applied to rescale the data. We attach a suffix of $ln$ to the rescaled feature to differentiate it from the original. For example, the rebalanced version of `session_min` is replaced with `session_min_ln`.

#### 5.2.2 Feature Reduction

| Rank | Feature | Extreme Value % |
|------|---------|-----------------|
| 1 | tx_rate_p75 | 24.82 |
| 2 | tx_rate_iqr | 24.82 |
| 3 | session_min | 13.38 |
| 4 | session_tx_min | 12.58 |
| 5 | tx_min | 12.51 |
| 6 | tx_rate_range80 | 12.17 |
| 7 | tx_rate_p90 | 12.17 |
| 8 | rx_rate_p75 | 11.84 |
| 9 | rx_rate_iqr | 11.81 |
| 10 | rx_max | 10.10 |

Table 4: Top 10 skewed features

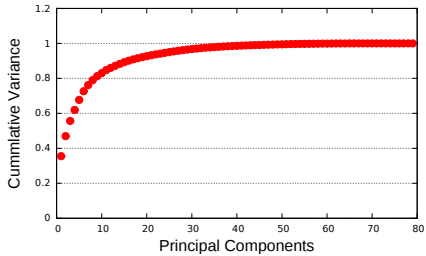

Figure 6: Variance captured by principal components.

| Coarse-Grain Classes | Fine-Grain Classes |
|----------------------|--------------------|
| rssi_ad | rssi_ad |
| rssi_range80 | rssi_range80 |
| rx_rate_p75_ln | rx_rate_p75_ln |
| tx_rate_p90_ln | tx_rate_p90_ln |
| session_rx_p10_ln | session_rx_p10_ln |
| rx_p25_ln | rx_p25_ln |
| rssi_min | rssi_p10 |
| session_rx_p25_ln | session_p75_ln |
| session_rx_p75_ln | session_max_ln |
| rx_p90_ln | tx_range80_ln |

Table 5: List of Top Features, w.r.t Coarse- and Fine-Grain Classes, Selected using CFS [11]

Classification performance is strongly related to the number of features employed. Looking over the set of features shown in Table 3, we may expect a certain level of correlation between some subsets. For example, it is very likely that the incoming traffic volume and outgoing traffic volume are likely to be highly correlated - since most client end-hosts (first) request content from remote servers which is subsequently delivered.

To quantify the extent of redundancy in our dataset, we carry out a Principal Component Analysis (PCA) [6] over the feature space. PCA is a well known linear transformation onto a new orthogonal space which has the property of concentrating the variance inherent in the original data. In a PCA transformed space, the principal components are rank ordered by the fraction of variance (in the original dataset) accounted for. Thus, the first principal component is associated with the largest amount of variance; each succeeding component has the highest remaining variance and under the constraint that it is orthogonal to all the previous components.

Figure 6 plots the cumulative variance contribution of the component axes obtained from a PCA transform. From the figure we see that *all* of the variance in the feature space can be accounted for with roughly 30 components, and about 20 components can explain over 95% of the variance in the data. This indicates that roughly half of the original features in our dataset are redundant and can be removed without adversely affecting classifier accuracy while at the same time speeding up the classifiers. However, one of the drawbacks of PCA is that the transformed space (where variance is concentrated along orthogonal axes) is not intuitive and does not inform us as to which features must be selected, and which can be dropped. What it does give us is a guideline for how many features should be retained.

Even though the feature space that we are dealing with is relatively modest, it is still an interesting excerise to try to reduce the set of features – it provides some insights into the most important features in the dataset. If this reduced set of features is small, it might even be feasible to perform the classification on the gateways themselves. In this paper, we examine the reduced feature set, and leave the viability of gateway-based classification to future work.

A number of techniques have been proposed in the literature towards feature sub-selection. We use a method known as *Correlation-based Feature Selection* (CFS) technique, that is known to produce feature subsets that generalize well to many different classification methods [11] and which produces a ranking over the features. CFS is based on the intuition that the ideal subset of features contain those features that are highly correlated with the labeled class, and at the same time poorly correlated with each other. Note that CFS is *supervised* in that it leverages the label to model feature dependence and thus, this part of our methodology is applied on the data (sub)sets $D_C$ and $D_F$. Examining the list of features selected with CFS (Table 5), RSSI-based features – particularly rssi_ad and rssi_auc) – and traffic rates (both overall and per session) are rated very highly by CFS.

## 6. RESULTS

In this section, we briefly describe two well known classification methods and subsequently present detailed results from applying these classifiers on the datasets $D_C$ and dataset $D_F$, and for which we use the labels inferred in Section 4 as the ground truth.

### 6.1 Classification Methods

In the course of working on this problem, we experimented with several different classification algorithms and techniques of varying complexity. In this paper, we describe the results from two particular methods – Decision Trees and Support Vector Machines (SVM). We found the SVM based classifier to have the best over-all accuracy. However, the results it provides are hard to interpret in terms of the features are are fed into it. On the other hand the Decision Tree based classifier yielded slightly lower accuracy, but it provides a framework to reason about the relevance of various features in classifying a device. We first report on the overall results, and later use the results of the Decision Tree to understand the importance of individual features.

In the earlier stage of our work, we extensively experimented with different classification algorithms/techniques i.e. advanced methods (Neural Networks, Support Vector Machines, Logistic Regression) as well as more traditional/simpler methods (Decision Tree, Naive Bayes and kNN) and at the end chose the following two classifiers (one from each group) based on their accuracy, speed, stability, and readability of results.

•**Decision Trees** are widely used in multi class classification applications. Construction starts with one node containing all the labeled instances and proceeds with a recursive splitting procedure. The splitting selects the most discriminative feature (using metrics such as Gini impurity, information gain, etc.) and partitions the parent node along the chosen feature. The recursion proceeds until no more partitioning is possible, and the tree is subsequently pruned (e.g. using minimal cost-complexity pruning). Weka implements a number of decision tree algorithms; we use SimpleCART [4] in our study. The defining characteristic of this method is that features are ranked and considered *sequentially*.

•**Support Vector Machines** (SVM, for short) are a family of powerful binary (supervised) classifiers which operate by identifying the boundary or *optimal separating hyperplane* between instances in different classes. One of the salient features of SVMs is the support for non-linear classification by mapping input data into a higher dimensional space (i.e., the so called "kernel-trick"). They computation in SVMs is to estimate the parameters of the hyperplane and a number of optimization methods have been proposed. In our work, we use a *linear* SVM with an implementation that uses the *sequential minimal optimization* (SMO) method [23]. One important advantage over Decision Trees is that SVMs can explicitly model dependencies *between features*.

**Parameter Tuning.** Both techniques described above involve a number of parameters (e.g., SVM penalty) and a key step in employing them effectively is to identify *good* parameter values. We use the Weka Machine Learning framework [10] which includes implementations for both of these, and which supplies a set of default parameters that are generally considered good. In our work, we further use a grid search over the available parameter space for each classifier and identify the optimal values that maximize classification accuracy.

We compare each classification method against each other, and against a baseline value that is obtained by a very trivial classifier, ZeroR, which essentially maps every device instance to the most common class in the training set. The result of the ZeroR classifier serves as a lower bound the accuracy.

## 6.2 Overall Results

For evaluation metric, we use the *accuracy* (defined as ratio of correctly-classified instances in the test set). Moreover, the results are based on 10-fold cross-validation where the data is divided into 10 equally-sized subsets and the model is trained 10 times, each time using 9 subsets for training and the remaining subset for test.The final accuracy is the average of these 10 runs. We also computed the F-measure metric for the classifiers and found the results to be consistent with accuracy (more precisely, in almost all cases F-measure is between 1 to 2 percent less than accuracy); we find the latter more intuitive to understand and only report on it.

Unless otherwise specified, the results in this section are shown as the *improvement* over the ZeroR classifier. We report this, and not the absolute value, for the following reasons. When the dataset is severely unbalanced (e.g. if a single device class dominates), the trivial classifier can perform as well as the more complex methods. Reporting the improvement over the trivial classifier – as we do – points to how much better a more principled classifier can do.

### 6.2.1 Accuracy

In Figures 7a and 7b, we show computed accuracy for the two classifiers and for three distinct feature sets. The baseline value, i.e., the starting value on the y-axis is the accuracy as reported by the trivial ZeroR classifier (recall that this corresponds to the population fraction of the most common device class). The feature subset *all* includes all of the features introduced in Table 3, while CFS:10 and CFS:20 correspond to the scenario where we only use the top 10 (or 20) features identified by the CFS algorithm. The accuracy reported corresponds to the optimal parameters (for each classifier) identified by the grid-search discussed previously. We point out that the improvement over using the default parameter settings in Weka are small (less than 2% at most).

Looking at Figure 7a in detail, i.e., accuracy in predicting coarse grained classes, we find that both classifiers perform quite well even as the SVM has slightly higher accuracy than the DecisionTree (90.47% as compared to 86.68%). Both classifiers do perform significantly better than the baseline. We suspect this is due to the fact that two categories are predominant in the device population and these are also easy to tell apart with the RSSI based features. Note that reducing the number of features does not necessarily degrade clas-sifier accuracy confirming the existence of redundancy in the feature as well as the effectiveness of the CFS method (DecisionTree's accuracy is generally less prone to feature reduction since the top-ranked features are less likely to be removed). However, we do point out that the number of features has a very large impact on the time taken to train the classifier and obtain a model.

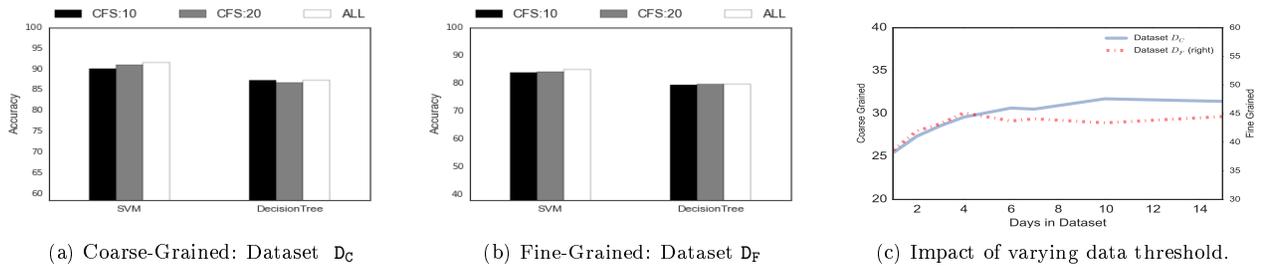

(a) Coarse-Grained: Dataset $D_C$  (b) Fine-Grained: Dataset $D_F$  (c) Impact of varying data threshold.

Figure 7: Classification Results

Turning now to Figure 7b which reports accuracy numbers when the classifiers are applied to the dataset $D_F$, we find that the accuracy drops by 6%-7%. As before the starting value on the y-axis is the accuracy reported by ZeroR. As expected this is lower than in the previous case – there are more device classes and the device population is more fragmented. Overall, we see that the SVM based classifier again outperforms DecisionTree (83.11% vs. 77.60%). In comparing the performance across the three different feature sets shown, we find mixed results across the classifiers. Overall, the reduction in performance due to using a smaller subset of features is not significant. The fact that the SVM based classifier consistently performs better than the Decision Tree (indeed, better than any other classifier that we experimented with) leads us to believe that there is some amount of coupling between features. Methods that examine features sequentially (e.g., Decision Trees) or those that assume independence (e.g., Naive Bayes) are unable to model this dependence and hence yield lower accuracies. However, the gains by using a more complicated method like an SVM are modest and this may be because some device classes are easier to identify than others.

Understanding this better requires looking at how well each classifier does in identifying individual device classes. Table 6 presents the confusion matrix that corresponds to the SVM classifier – we picked the best performing technique – applied to fine grained data (dataset $D_F$). We omit a discussion of the same for coarse grained labels due to a lack of space. In Table 6, the columns are labeled by the letters *a* through *g*, which correspond to the classes indicated on the rows. We see two patterns emerge from the matrix. First, the classifier often labels smartphones as tablets,[2] and vice versa. Second, the classifier also has difficulty clearly distinguishing PC devices (laptops and desktops) from network extenders. To rationalize the former, one possible reason is they really are used in similar ways (note that we do not have any information about the *cellular* activity of the smartphone). It is entirely possible that a particular individual would use his smartphone to watch lots of media, and this is what another individual uses his tablet device for. Looking at the second trend, i.e., the confusion between compute devices and network extenders, consider that a network extender does not really have a *personality* of its own, and simply aggregates many devices behind it. Thus, when the traffic features do match up between these classes (the distributions for these two entities do overlap as seen in Fig. 2), they would appear indistinguishable to the gateway. Furthermore, we manually inspected the network extenders that are incorrectly labeled as laptops/desktops and noticed that in most, there are a small number of hostnames associated with the MAC address (indicating few devices behind the extender).

---
[2] This confusion does not impact the accuracy for the coarse

| Classified As → Actual Label ↓ | (a) | (b) | (c) | (d) | (e) | (f) | (g) |
|---|---|---|---|---|---|---|---|
| Smartphone (a) | **396** | 1 | 8 | 13 | 2 | 1 | 1 |
| Network/WiFi Ext. (b) | 3 | **76** | 37 | 1 | 0 | 1 | 1 |
| Laptop/Desktop (c) | 13 | 6 | **374** | 1 | 0 | 1 | 0 |
| Tablet (d) | 38 | 2 | 7 | **43** | 1 | 0 | 1 |
| Game Console (e) | 13 | 2 | 5 | 3 | **11** | 1 | 0 |
| Printer/Scanner (f) | 4 | 1 | 2 | 0 | 2 | **2** | 1 |
| OTT box (g) | 0 | 3 | 5 | 2 | 0 | 1 | **3** |

Table 6: Confusion Matrix: SVM applied to dataset $D_F$

### 6.2.2 Data Sufficiency

Most of our features are constructed starting from a time series of a particular metric. It is then interesting to ask if there is a dependence between observation time, i.e., the period of time over which the time series is constructed, and the resulting accuracy in classifying the device. We reformulate this question to ask: *what is the minimum time duration for a device to be observed before it is can be classifed to a degree of accuracy?*

---
grained classes since both of these are grouped as Mobile Handheld devices

To consider an extreme example where we use a threshold of 1 day. This implies that a device that has been active (observed by the gateway) for at least 1 day is included into the training set. Thus, with a threshold of 1 day, the training set may include devices active for just a day or slightly longer. Obviously, the features extracted from these devices have a lot of variance and are likely to be misclassified. At the other extreme, if we were to use a high threshold (say 1 month), we would reject devices that were active for less than that period. This reduces the training set and rejects devices that may be used infrequently.

To find the right balance and identify the appropriate cut off duration, we ran a set of experiments where the *threshold* (number of days a device was active) was varied. The results for the SVM classifier are shown in Figure 7c. The x-axis indicates the threshold used (number of days), and the y-axis describes the relative improvement over the ZeroR classifier. For e.g., when the threshold is 1 day, the accuracy of the SVM classifier is 25% more than the ZeroR classifier. As seen in the figure, there is a initial improvement in the accuracy (around x=4) after which it begins to flatten out. As x increases, two things happen: (i) more devices are rejected from the training set, and (ii) this also changes the device class distribution. The rejected devices are likely to have been misclassified (so removing them increases accuracy). Further, as the class distribution changes, the baseline accuracy also changes, which further amplifies the difference between the ZeroR classifier and the SVM. The surprising result (for us) was that a device has to be active for only a few days for it to be classified accurately. Hence, as a tradeoff between maximum accuracy and maximum dataset size, we have used the threshold value of 3 in this paper to define the transient versus non-transient devices (explained in Section 1).

### 6.3 Incorporating MAC Information

Recall from Section 3.1 that the MAC OUI identifies the manufacturer of the network interface. Manufacturers vary greatly in the number of *types* of devices that they manufacture. On one hand, `Apple` builds devices that cover *all* of the coarse grained device classes (and here the OUI is not likely to possess significant discriminatory power). On the other hand, `Roku` manufactures exactly one type of device – an OTT box; here, the OUI completely determines the class of device. Most manufacturers fall somewhere inside this spectrum and this can be leveraged to improve the classification accuracy.

It is very important to point out that the correlation between OUI to category correlation is not a very general one. Conceivably, a different dataset has contains predominantly Apple manufactured devicesis unlikely to benefit from the addition of the OUI as a feature.

Since the learned classifier function(s) depend greatly on the distribution of device manufacturers across devices – and this can vary a great deal based on country, user demographic, etc. – we do not incorporate the OUI into our core set of features. That being said, examining its utility inside our specific dataset is interesting.

Figures 8 shows the *improvement* in classification accuracy for the datasets $D_C$ and $D_F$ by incorporating the OUI as categorical feature. Note that we used the entire set of features here (rather than the subsets extracted from CFS). The figure clearly shows that the classifier accuracy increases (by more than 10% in the best case). The improvement is greater for classifiers operating on the fine grained data. To point to an example: the accuracy of the SVM *increases* by about 7% when applied to *fine grained dataset*. However, the higher gain in accuracy corresponds to the DecisionTree classifier. Recall that this classifier recursively splits the data based on the feature that has the highest (remaining) discriminative power; the OUI is selected early because of the heterogeneity of OUI's in our dataset. This factor also cautions against generalizing these results too much; a different dataset with a different distribution of manufacturers may produce completely different results.

### 6.4 Human-Readable Behavior Inference

While the SVM based classifier is the best performing, the results are hard to interpret and understand which particular features lead to certain devices being classified as they are, or their relative importance in deciding on the final labeling. On the other hand, the DecisionTree proceeds by greedily partitioning the dataset by successively selecting discriminating features. Thus, the partitioning order directly indicates the relative importance of the associated feature. Note that the individuals labels generated by DecisionTree might be different from those assigned by SVM; however the performance of the DecisionTree is not very far from the SVM. Fig. 9 shows a tree generated using SimpleCart. Here, features (internal nodes) are ordered by the distance to the root of the tree. In this particular case, the most important feature is determined to be *rssi_ad* (this captures device mobility). Notice that the right side of the tree, where the feature value is greater than 2.04 only contains leaf nodes relating to Mobile hand held devices. Knowing that a device has high mobility is sufficient to rule out devices like computers, game consoles, and OTT boxes, which are typically used in a few fixed locations. The structure of the tree also provides a few hints that could explain the *confusion* between smartphones and tablets as well as PCs and Network Extenders (cf. Table 6); these appear as siblings in the upper levels of the tree (near the root).

Going further, we seek to understand the *essence* of a particular device class and to be able to describe it succinctly with a set of concise rules. To state it another way, we wish to uncover what makes a smartphone a *smartphone*, and not a game console. There are two drawbacks in directly using the decision tree as constructed previously: (i) a certain class may appear in several leaf nodes throughout the tree, which confounds efforts to describe the classes purely in terms of a single path from the root of the tree, (ii) more importantly, minority classes are not captured in the tree (in our tree, for example Printer/Scanner), as the classifier can "afford" to ignore them (labeling them incorrectly degrades accuracy very little). To address this, we employ the *one-versus-all-the-rest* approach for individual classes and use the ConjunctiveRule [25] classifier implemented in Weka; this classifier learns a simple set of conjunctions that can be used to predict nominal class labels. Before we apply it, we need to balance the device classes and to this end we use a technique called Synthetic Minority Oversampling TEchnique (SMOTE [8]). This generates new artificial instances for the minority class to make the dataset more balanced. The result of the ConjunctiveRule learner is described in Table 7, which roughly correspond to concise behavior descriptors of the device. The table also indicates the accuracy that results from using *only* the induced rule to differentiate devices of that class from the rest of the population. For some fine grained categories, single features are extremely discriminative; for e.g., `rssi_ad` (almost) completely describes smartphones, and thresholding `tx_p25_ln` completely captures printer/scanner devices. The latter is rule is quite intuitive since we expect these devices to be used very rarely, and thus, would generate low volumes of traffic on an average day. One salient property of these rules is that they can be used directly on the gateways to carry out a reasonably good classification of the devices in the home.

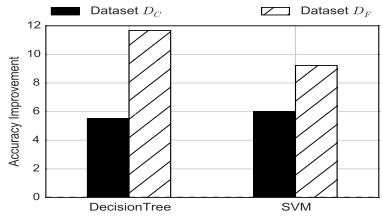

Figure 8: Impact of incorporating MAC OUI

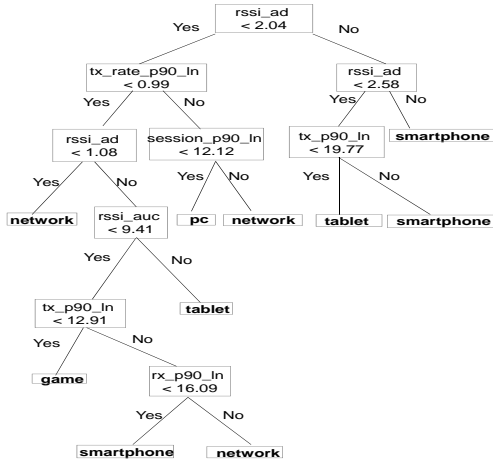

Figure 9: Output of SimpleCART on dataset $D_F$

| Category | Conj. Rule | Accuracy |
|---|---|---|
| Smartphone | `rssi_ad > 2.1` | 89.6 |
| Printer/Scanner | `tx_p25_ln < 12.71` | 88.76 |
| Laptop/Desktop | `tx_rate_p90_ln > 0.70` | 86.13 |
| Game Console | `rx_med_ln < 12.5 AND tx_rate_max < 14633.1` | 83.76 |
| OTT | `rx_rate_max < 800` | 81.61 |
| Network Eqpmt. | `session_range80_ln > 12.1 AND session_rx_rate_p10_ln > 1.1 AND traffic_days > 25.51` | 76.01 |
| Tablet | `rssi_ad > 1.2 OR session_rx_rate_p75_ln < 4.1)` | 69.84 |

Table 7: Results of Single Conjunctive Rule Learner

## 7. CONCLUSIONS

In this paper, we described a methodology to identify the category of a networked device based on on low level indicators of network activity logged on home internet gateways. We analyzed in detail a dataset of 240 subscriber homes and extracted a number of features that succinctly capture the traffic and spatial behavior of devices in our dataset. We also defined a two level taxonomy of device categories (coarse and fine) and used a number of heuristics on static device descriptors to associate devices with these labels (which were then extensively checked by manual inspection). We then experimented with a number of well known classification methods towards predicting the labels that we previously obtained. Overall, we find that the coarse grained (higher level) label of a class can be inferred with very high accuracy (91% with the best case SVM classifier), with fine grained labels accurate up to 84%. We also examined the impact of incorporating mac address information into the classification and showed an accuracy improvement of about 6% (coarse grained) and 8% (fine grained). While these improvements may not generalize outside of our specific dataset, they provide useful insights. Moving past just classifier performance, we attempt to understand the relationship between feature types and individual device classes. To this end, we construct a set of simple, concise predicate conjunctions (over the features) that capture the latent character of each device type. One of the takeaways in our work is that the high accuracy is realized even with a relatively small feature set, and all of which focus purely on two aspects of device behavior – how much traffic it exchanges on the network, and its positioning (relative to the gateway) over time. This was enabled by a detailed dataset exploration and selection of features that combined intuition with existing best practices. To highlight one particular case of this: all of the spatial features rely on the same metric reported from the gateway; by extracting various behaviors from this single metric, we are able to successfully discriminate between different kinds of mobile devices.

Our work is described in the context of an ISP (which owns and operates the home internet gateway) that wishes to classify devices in a customer's home. The model that we envision is that there is an initial offline procedure (with a large training set) carried out in the ISP's premises to learn the classification models. Subsequently, gateways periodically report the feature summaries the ISP where they are run against the selected model(s) after which device labels are sent back to the gateway. In the future, we plan to investigate classification methods that can directly be run on the gateway itself.